# Characterizing Transition-Metal Dichalcogenide Thin-Films using Hyperspectral Imaging and Machine Learning

Brian Shevitski[1,2,3,4], Christopher T. Chen[4], Christoph Kastl[4,5], Tevye Kuykendall[4], Adam Schwartzberg[4], Shaul Aloni[4*] and Alex Zettl[1,2,3*]

[1]Department of Physics, University of California at Berkeley, Berkeley, CA 94720, U.S.A.

[2]Materials Sciences Division, Lawrence Berkeley National Laboratory, Berkeley, CA 94720, U.S.A.

[3]Kavli NanoEnergy Sciences Institute at the University of California at Berkeley and the Lawrence Berkeley National Laboratory, Berkeley, CA 94729, U.S.A.

[4]The Molecular Foundry, Lawrence Berkeley National Laboratory, Berkeley, CA 94720, U.S.A.

[5]Walter-Schottky-Institute and Physik Department, Technical University of Munich, Garching, 85748, Germany

*To whom correspondence should be addressed: azettl@berkeley.edu, saloni@lbl.gov

**Abstract**

*Atomically thin polycrystalline transition-metal dichalcogenides (TMDs) are relevant to both fundamental science investigation and applications. TMD thin-films present uniquely difficult challenges to effective nanoscale crystalline characterization. Here we present a method to quickly characterize the nanocrystalline grain structure and texture of monolayer $WS_2$ films using scanning nanobeam electron diffraction coupled with multivariate statistical analysis of the resulting data. Our analysis pipeline is highly generalizable and is a useful alternative to the time consuming, complex, and system-dependent methodology traditionally used to analyze spatially resolved electron diffraction measurements.*

Transition-metal dichalcogenides (TMDs) display emergent properties when reduced to single, two-dimensional (2D) layers. A transition from indirect to direct band gap[1,2], the emergence of charge density waves[3,4] an increase in mobility[5–7], and the presence of valley polarization[8–10] are a few of the important properties that are manifested in the monolayer limit.

Polycrystalline TMD thin films can be grown at wafer scale and lend themselves to scalability[11,12]. These films have a high density of intrinsic grain boundaries and other defects that can influence physical properties and drive exotic correlated electron effects and emergent phenomena[4]. In this communication, we characterize large area polycrystalline thin-films of $WS_2$ using scanning nanobeam electron diffraction, also called four-dimensional scanning transmission electron microscopy (4DSTEM) to identify the local crystalline texture and structure. We employ advanced multivariate statistical analysis (MVA) techniques to rapidly extract pertinent information, namely the grain structure of the $WS_2$ films, from the complex, high-dimensional 4DSTEM data.

$WS_2$ films are grown directly on electron transparent SiN membranes, resting on Si supports, using a previously described technique[13]. Samples are prepared by depositing a coating of 10 nm of $SiO_2$ on the SiN membrane, as well as the back and edges of the support window using plasma-enhanced atomic layer deposition (PE-ALD). This provides an ideal growth substrate on the electron transparent window and protects the Si support frame from chemical conversion during subsequent steps. 2 nm of $WO_3$ is deposited onto the substrates using PE-ALD. The metal-oxide precursor is converted to $WS_2$ in a dry (< 10 ppm water) tube furnace at 800 C using $H_2S$ as a chalcogenization agent.

In a 4DSTEM experiment (Fig. 1(a)), we acquire diffraction data over a wide area of the sample. This is in contrast to traditional dark-field (DF) TEM imaging, where a physical aperture is placed in the diffraction plane of the instrument at the location of a Bragg spot, resulting in an image formed by Bragg scattered electrons that have passed through the aperture. DF-TEM characterization uses a series of aperture images, acquired at several aperture positions, to construct a map of the spatial distribution of the crystalline grains in a sample[14,15]. In contrast, 4DSTEM simultaneously acquires all possible aperture positions, including those that do not fall directly on a Bragg peak.

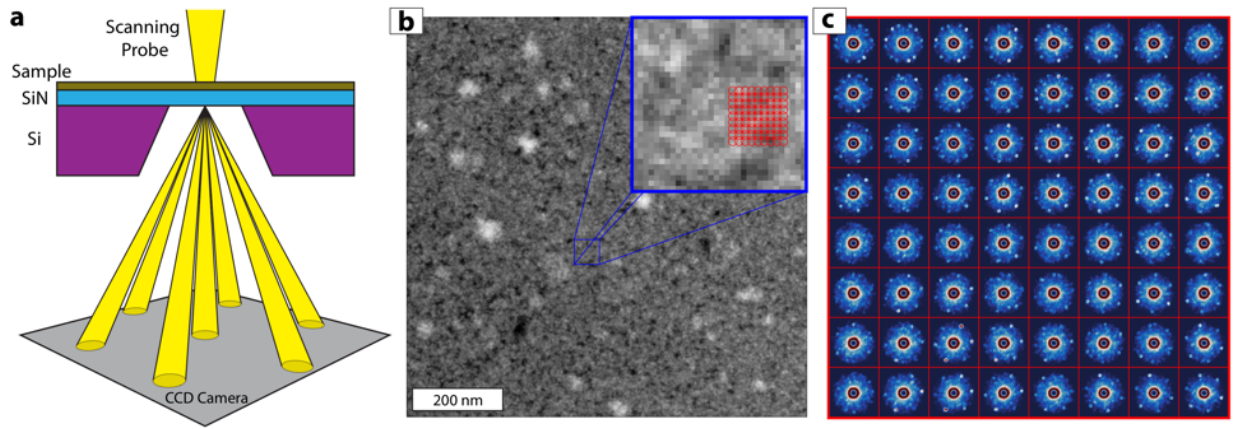

Fig. 1: (a) In a 4DSTEM experiment, a small convergence angle probe of high-energy electrons is rastered across a thin sample suspended across a supporting frame (typically with a thin, electron transparent window). A CCD camera in the back focal plane records the electron count (reflecting the diffraction pattern) at each probe location, thus measuring the local crystallography of the sample with nanometer scale resolution. The films used in this study are synthesized by converting $WO_x$ films deposited directly onto SiN TEM membranes. (b) Annular dark-field (ADF) STEM image of a $WS_2$ film. The blue box (inset) is a zoom-in (7x magnification) for the small blue square area from the center of the image. (c) 64 representative diffraction patterns acquired from the $WS_2$ sample at the spatial locations indicated by the red grid in (b) The diameter of the red circles in (b) indicate the approximate probe size. In (c), each red box represents a spatial pixel of size 2 nm sampled by a 2.7 nm probe. The field of view of the diffraction patterns in the red boxes is 10.8 $nm^{-1}$.

Fig. 1(b) shows a conventional STEM image of a $WS_2$ thin-film acquired using an annular dark field (ADF) detector. The contrast in this image indicates differences in thickness, mass density, and local crystallography of the sample. The bright regions are the thin film, the dark regions are voids, and the very bright spots are regions of contamination. The sample presumably has a distribution of grain sizes and orientations, but this is not directly apparent from the STEM image of Fig. 1(b).

We now investigate the same $WS_2$ film via 4DSTEM. The red circles in the inset of Fig. 1(b) correspond to approximate positions of the probe during the 4DSTEM mapping. Fig. 1(c) shows a visualization of the associated raw diffraction pattern data. Each red box in Fig. 1(c) presents spectral data collected from the corresponding spatial pixel (red circles) in Fig. 1(b). It is apparent that each diffraction pattern has a mixture of two main features: sharp, bright spots arranged in an approximately hexagonal pattern arising from Bragg scattering from the crystalline planes of the thin-film and a diffuse component with approximate azimuthal symmetry that arises from the amorphous support substrate (the highly saturated central spot from unscattered electrons contains no useful information and has been masked in Fig. 1(c)).

The data in Fig. 1(c) hint at differently oriented crystallites (i.e. domains) with hexagonal symmetry within the sample. However, the true rotational symmetry and detailed domain structure are not easily individually identified and assigned by eye. In fact, as we find below, the rotational symmetry in this specimen is not six-fold at all. This illustrates the general difficulty of directly visualizing or assigning unambiguous meaning to higher dimensional data, particularly 4DSTEM data sets.

Fig. 2 shows the results of a traditional analysis of the 4DSTEM data. Bragg peaks are detected in 10 randomly chosen diffraction patterns using difference of Gaussian (DoG) blob detection. The detected blobs are extracted and averaged together to create an exemplar for the diffraction spots, which is then used as a template. Bragg peaks are detected in each diffraction pattern of the 4DSTEM data using cross-correlation matching of the template. This preliminary set of diffraction peaks is enhanced by removing any matches that fall outside of a well-defined range of reciprocal space radii (3.43 nm$^{-1}$ ≤ q ≤ 3.94 nm$^{-1}$), corresponding to the in-plane reciprocal lattice constant of WS$_2$ ($q_0$ = 3.67 nm$^{-1}$). The image shown in Fig. 2 is generated by drawing lines corresponding to the orientations of all Bragg peaks at each spatial pixel. The color scale indicates the angle, in degrees, of each Bragg reflection (modulo 60 degrees).

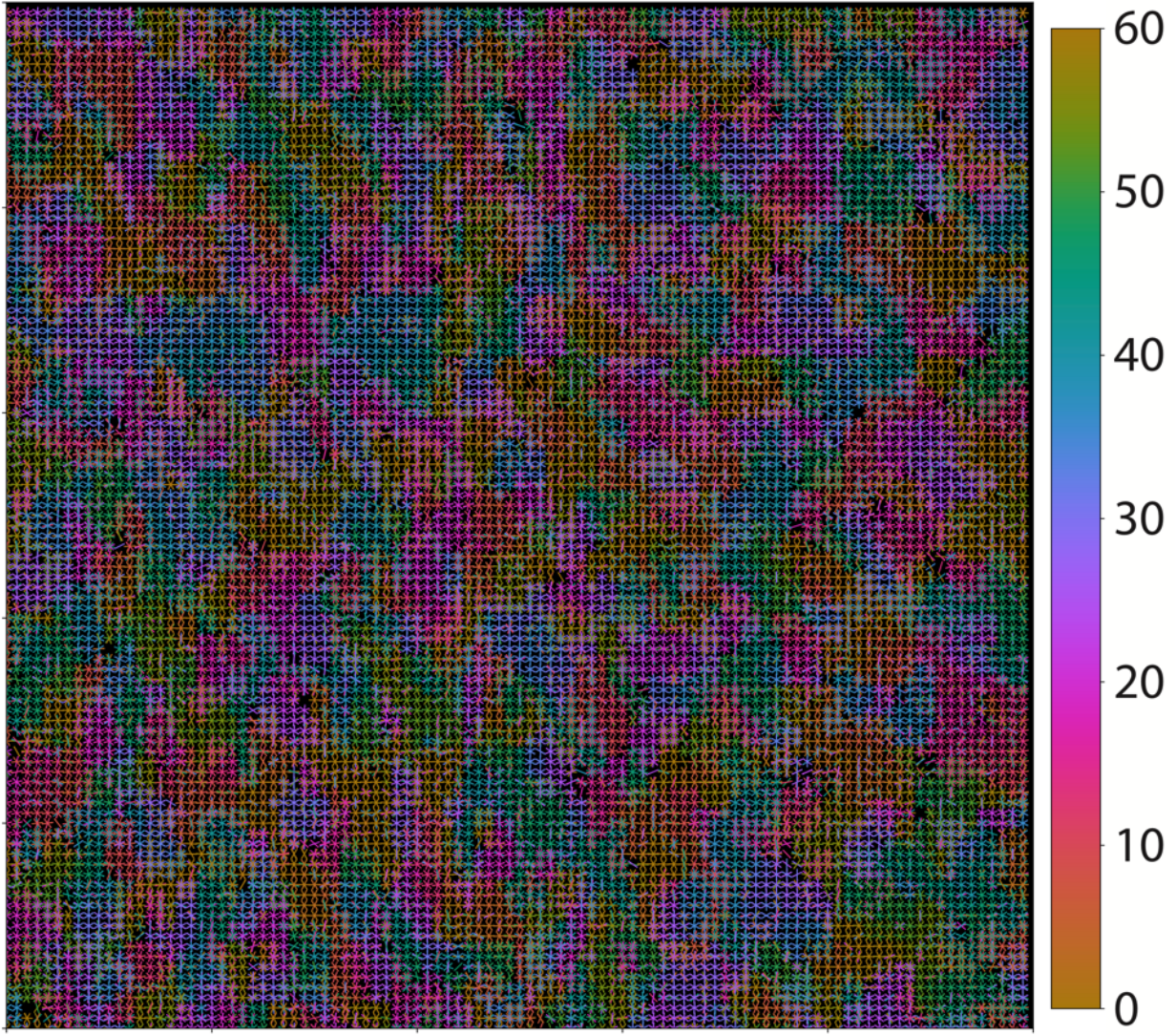

Fig. 2: Visualization of grain structure in a WS$_2$ thin film. Peaks are detected in the diffraction pattern acquired at each spatial location of the 4DSTEM data. The angle of each peak is extracted and used to generate the colored lines shown in the figure. The color scale indicates the angles of each line, modulo 60 degrees. The image field of view is 200 nm.

Fig. 2 indicates that the specimen is comprised of many small grains with lateral sizes on the order of ten nanometers. The majority of the domains do not overlap, but there are regions on the sample with multiple crystallographic orientations at a single spatial location. The result presented in Fig. 2, while striking, is time consuming to construct and the data analysis relies on *a priori* knowledge of the crystal structure, implying that the methodology is not necessarily generalizable to other systems. Furthermore, even though the image presented in Fig. 2 has a significant reduction in size and dimension compared to the original data set, there is still too much information density to allow the facile extraction of the most relevant properties of the system under investigation (e.g. the precise distribution of grain sizes and orientations).

MVA techniques are extremely useful for tackling the problems of dimensional reduction and information extraction from complex sets[16–21]. These statistical techniques result in a simplified representation of high-dimensional data consisting of a small number of low dimensional components which convey the general trends in the original data set. Some preliminary attempts have been made to approach 4DSTEM analysis using MVA, with varying degrees of success[21,22]. With this in mind, we apply MVA methodology, using the approach outlined in Fig. 3. Prior to MVA decomposition, the data are pre-treated. Many MVA techniques are highly sensitive to small shifts and outliers in the data, which can either be a blessing or a curse. In order to minimize artefacts in the MVA output, the data are first aligned (there are small shifts in the diffraction patterns recorded at different spatial pixels) and outliers are removed (e.g. cosmic rays result in "hot pixels" and the intensity from the central beam is both highly saturated and much greater than the intensity from scattered electrons).

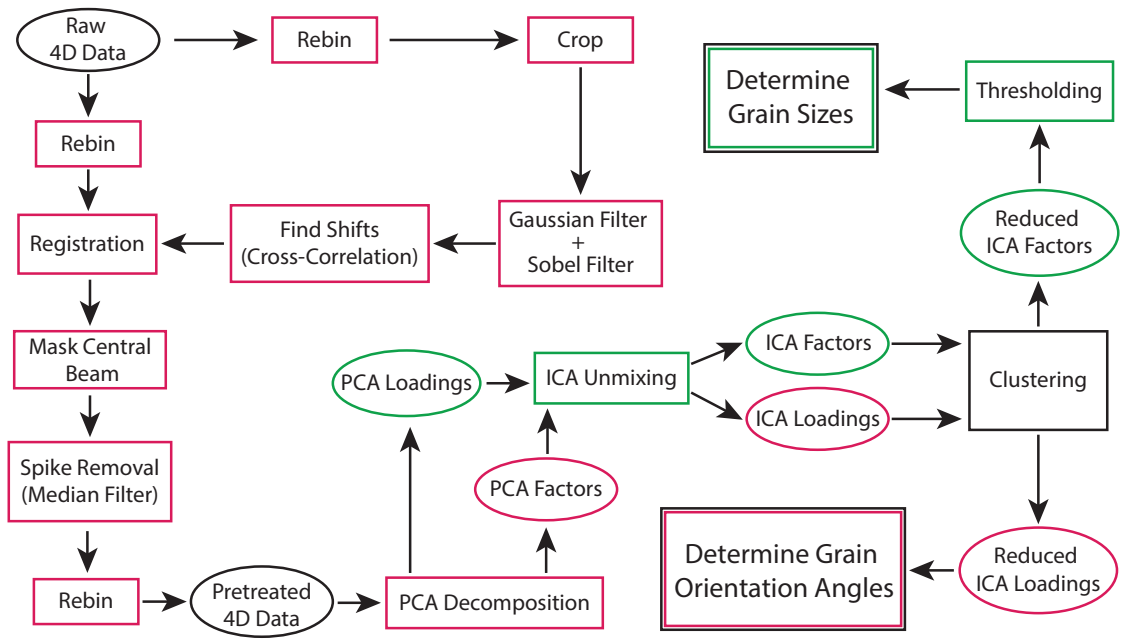

Fig. 3: Flowchart for MVA workflow of 4DSTEM data. Red items denote spectral pixels and green denote spatial pixels. Rectangular boxes are operations and ovals are data.

The shifts between diffraction patterns are calculated by cross-correlation of the central beam followed by interpolation of the data to bring it into registration. The cross-correlation is enhanced by applying a noise reducing Gaussian filter followed by an edge-finding Sobel filter to the data before calculation of the cross-correlation coefficient[23].

Hot pixels are removed using a 3x3 median filter and the intensity distribution from the central beam is masked with a circular disk. The final step of data pre-treatment is rebinning each diffraction pattern, making the size of the data more manageable and reducing the computation time for subsequent steps.

The first step of the MVA portion of the data analysis workflow is to decompose the data into a new basis using principal component analysis (PCA)[17,24]. The Principal Components (PCs) are orthogonal and describe which parts of the data contain the most variance. The first PC accounts for the most variance in the data, the second PC accounts for the second most variance, and so forth. The PC basis is constructed from the data covariance matrix, $\mathbf{C}(\mathbf{k})$, given by:

$$\mathbf{C}(\mathbf{k}) = \sum_{x}(\mathbf{D}(\mathbf{x},\mathbf{k}) - \overline{\mathbf{D}}(\mathbf{k}))(\mathbf{D}(\mathbf{x},\mathbf{k}) - \overline{\mathbf{D}}(\mathbf{k}))^{\mathbf{T}} \qquad (1)$$

$\mathbf{D}(\mathbf{x}, \mathbf{k})$, the as-acquired data set, is a function of two spatial directions ($\mathbf{x}$) and two spectral dimensions ($\mathbf{k}$), $\overline{\mathbf{D}}(\mathbf{k})$ is the mean over the spatial dimensions, and $\mathbf{T}$ denotes the matrix transpose. The PC basis vectors, $P_\alpha(\mathbf{k})$, are the eigenvectors of $\mathbf{C}(\mathbf{k})$. In the PC basis, the data are represented as:

$$\mathbf{D}(\mathbf{x}, \mathbf{k}) = \sum_{\alpha=1}^{N} a_\alpha(\mathbf{x}) P_\alpha(\mathbf{k}) \qquad (2)$$

where $a_\alpha(x)$ are the spatially varying weight coefficients and N is the dimension of the raw data. N is either the total number of spatial pixels or total number of spectral pixels, whichever is smaller. Traditionally, the weights (which, for the data discussed here, are real-space images) are called the PC loadings, and the PCs (which, for the data discussed here, are diffraction patterns) are called the PC factors.

Fig. 4 shows the primary features of the PCA decomposition. The first several components (1-5) are the most important and they indicate clear spatial structure. We observe a rotationally symmetric component, related to the mean response of the sample, as well as azimuthally varying ring-shaped components which describe the intensity of the Bragg spots throughout the sample. The next components (15-19) have less clear spatial structure, but decidedly more complex spectral structure; these components describe complicated intensity variations of the diffuse background. Components 50-54 have no discernable spatial structure and an intricate spectral structure that has no immediately clear meaning. The final components (500-504) have no structure either spatially or spectrally and show the descent of the components into random noise.

A fundamental assumption of PCA is that a data set can be described to a high degree of precision by retaining only $N' \ll N$ components. It is assumed that the most important parts of the data (those with the highest variance) reside in the earlier components, while the later components contain primarily high-frequency noise, similar to Fourier decomposition and compression; inspection of Fig. 4 suggests that this assumption is valid. In this case, the reconstructed model of the data, $\mathbf{M}_{PCA}$, is given by:

$$\mathbf{M}_{PCA}(\mathbf{x}, \mathbf{k}; N') = \sum_{\alpha=1}^{N'} a_\alpha(\mathbf{x}) P_\alpha(\mathbf{k}) \qquad (3)$$

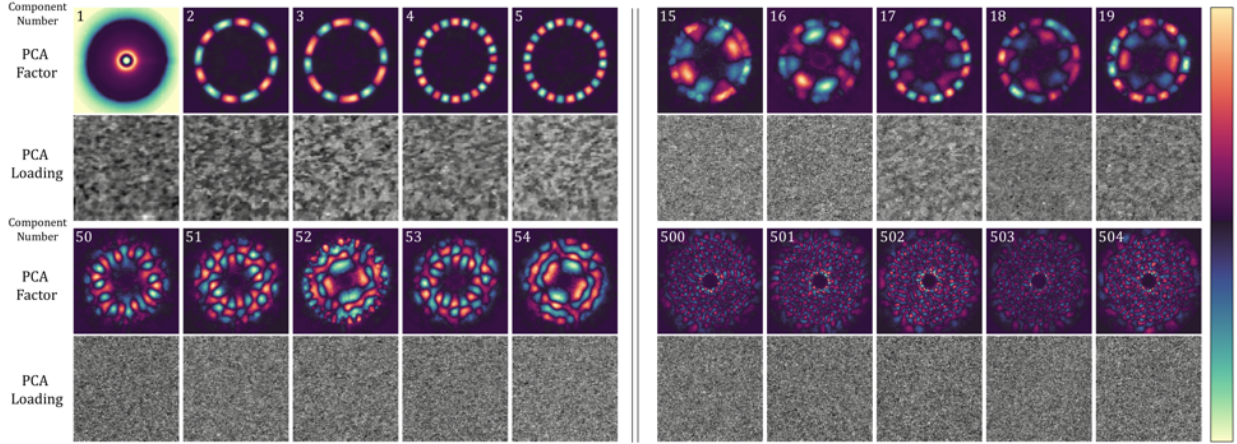

Fig. 4: Results of principal component analysis (PCA) for polycrystalline $WS_2$. Selected k-space PCA factors (analogous to diffraction patterns) above their associated real-space PCA loadings (analogous to images) are shown above. Similar to Fourier decomposition, the data are expressed as a linear combination of a new basis. The information basis describes the variance of the data. The first member describes the highest variance, the second member the second highest variance, and so on. As the basis component number increases, the k-space PCA factors and associated real-space PCA loadings move from describing the intensity of the primary diffraction spacing (components 1-19), to describing the intensity of scattering from the amorphous support substrate (components 15-54), to finally the noise of the detector (components 500-504). The field of view for each PCA factor and loading is 10.8 nm$^{-1}$ and 200 nm, respectively.

The PCA components are orthogonal and thus do not necessarily describe physical processes. In order to decompose the PCA components into a new basis that more accurately reflects the physical reality of the sample, we employ independent component analysis (ICA) unmixing to perform blind source separation (BSS) of the spatial PCA loadings[25–27].

BSS assumes the conjecture that if signals are from distinct physical processes, those signals will be statistically independent. The crux of the method is the reasonable (but logically unwarranted) assumption that this conjecture can be reversed; namely, BSS assumes that if signals are statistically independent, then they originate from different physical processes.

Two signals, X and Y, are uncorrelated if $\langle XY \rangle = \langle X \rangle \langle Y \rangle$ where the brackets denote the expectation value. Two signals are statistically independent if $\langle X^p Y^q \rangle = \langle X^p \rangle \langle Y^q \rangle$ for all positive integers p and q. Statistical independence is related to correlation but is a stronger condition. For example, the x and y coordinates of a body in uniform circular motion are uncorrelated but not statistically independent.

The goal of ICA is to un-mix a set of components into a new basis that has maximal statistical independence. We use a subset of the PCA loadings (real space images) as the set of components for separation. In ICA, the model of the data is given by:

$$\mathbf{M}_{ICA}(\mathbf{x}, \mathbf{k}; N'') = \sum_{\alpha=1}^{N''} c_\alpha(\mathbf{k}) I_\alpha(\mathbf{x}) \qquad (4)$$

where $\mathbf{M}_{ICA}$ is the new ICA model which is again a function of two spatial directions $\mathbf{x}$ and two spectral dimensions $\mathbf{k}$, $c_\alpha(\mathbf{k})$ are the spectrally varying weight coefficients, $I_\alpha(\mathbf{x})$ are the spatially varying independent component maps, and $N''$ is the reduced dimension of the independent component space. Each new independent component is constructed as a linear combination of principal components,

$$I_\alpha(\mathbf{x}) = \sum_{\beta=1}^{N'} w_{\alpha\beta} \, a_\beta(\mathbf{x}) \qquad (5)$$

where $a_\beta(\mathbf{x})$ are the principal component loadings and $w_{\alpha\beta}$ are entries of the mixing matrix. The FastICA algorithm[28] is a reliable method to efficiently determine the mixing matrix, which gives the set of independent components that have maximal statistical independence from one another.

After the mixing matrix has been computed using FastICA, the complementary k-space independent components $c_\alpha(\mathbf{k})$ are determined by:

$$c_\alpha(\mathbf{k}) = \sum_{\beta=1}^{N'} w_{\alpha\beta} P_\beta(\mathbf{k}) \qquad (6)$$

The independent components of Eqs. (5) and (6) underpin the model that emerges from the original experimental data, and as such provide a method to rapidly examine the major features of a large, high-dimensional data set.

One of the biggest challenges to utilizing MVA is the determination of the number of components to keep for the final reconstruction (N′). Fig. 5 outlines several metrics for selection of N′. The PCA Scree plot (Fig. 5(a)) shows what proportion of the total variance each component adds to the data. Inspection of Fig. 5(a) reveals that there are four distinct regimes of components, denoted by the vertical colored lines. The first component alone accounts for nearly 95% of the total variance in the data (left of the red line). The curve defined by the components in the second regime (left of the green line) has a distinct shape, and each component accounts for between ~ 0.1% - 1% of the variance in the data. The remaining two regimes (left and right of the blue line) appear as a smooth curves with an elbow around 500 components. These values (1, 10 and 500 components) are useful trial values of N′ for inspecting general trends in the PCA decomposition.

The integral of the Scree plot (Fig. 5(b)) shows how much each component adds to the cumulative variance, and is a useful metric for determining the final number of components in a decomposition. A common method for choosing N′ is to keep all components below an arbitrary threshold in the cumulative variance plot or the Scree plot. Fig. 5(b) shows three different choices of cutoff, the knee of the curve (red), 99% of the total variance (green), and 99.9% of the total variance (blue). In this plot, the vertical lines indicate the final output dimension and the horizontal lines indicate the choice of threshold.

The mean square reconstruction error (MSE) is given by:

$$e_M^2(N′) = \frac{1}{N_x N_k} \sum_{x,k} |D(x,k) - M_{PCA}(x,k;N′)|^2 = \frac{1}{N_k} \sum_{\alpha=N′+1}^{N} \lambda_\alpha \qquad (7)$$

Here, $\lambda_\alpha$ are the eigenvalues of the data covariance matrix ($C$), $N_x$ and $N_k$ are the number of spatial and spectral pixels, $D$ is the data set, and $M_{PCA}$ is the reconstructed PCA model. The last expression in Eqn. (7) can be quickly calculated for all values of N′ and can be used to rapidly determine a value of N′ based on the MSE. PCA decomposition can be used for lossless data compression and storage when N′ is chosen such that the root mean square reconstruction error (RMSE) is equal to the noise floor of the measurement[17]. The resulting representation of the data has increased signal to noise and decreased data size with the same information content. Fig. 5(c) shows the RMSE as a function of N′, calculated using equation (7).

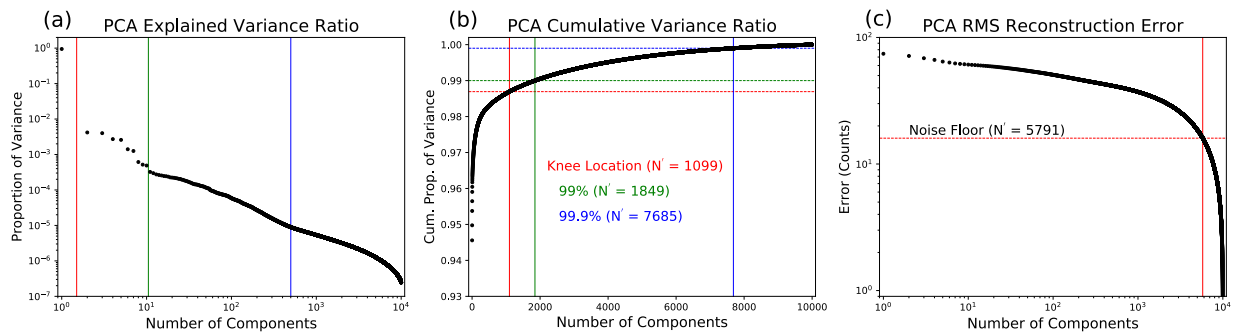

Fig. 5: Determination of number of PCA components using Scree plot and reconstruction error. The internal structure of the PCA Scree plot, shown in panel (a), shows four regimes of components, delineated by the red, green, and blue vertical lines. The colored lines in the cumulative PCA scree plot in panel (b) show three (arbitrary) cutoffs for dimensional reduction, listed on the figure. Panel (c) shows the RMS reconstruction error, along with the noise floor of the detector.

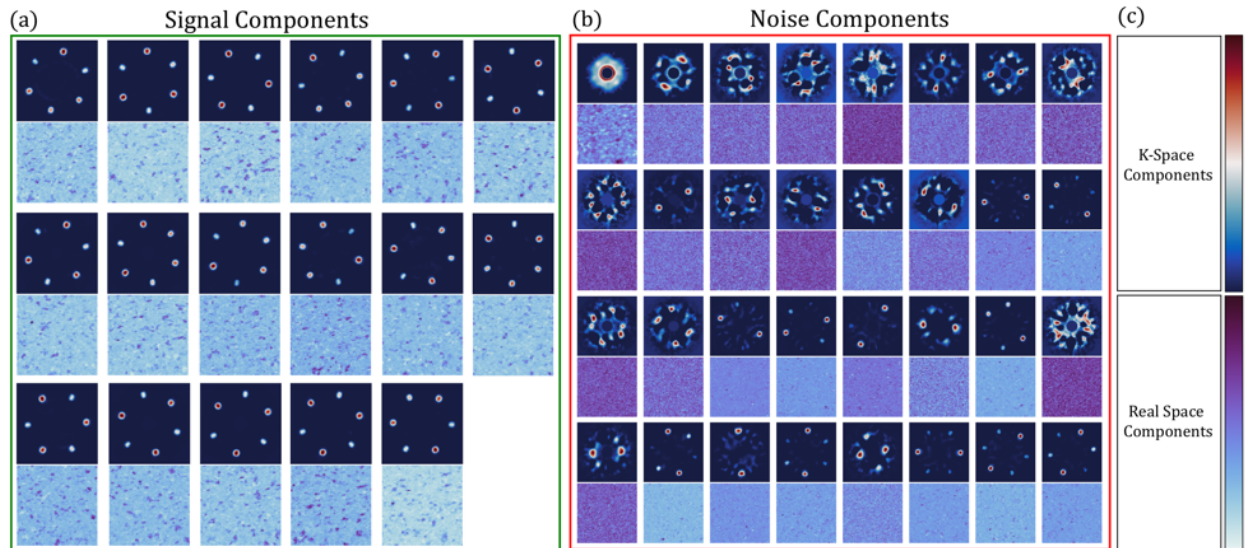

Fig. 6: ICA unmixing of PCA components. The k-space components (top sub-panels) and real-space components (bottom sub-panels) are sorted into two categories of noise or signal. The signal components in panel (a) appear as three-fold symmetric diffraction patterns in k-space and a collection of small crystalline grains in real-space. The color scales for the real-space and k-space components are shown in (c).

We use intuition gained from the PCA Scree plot to choose the number of components for the ICA input (N′) and output (N″). Fig. 6 shows the results of ICA unmixing using 498 PCA inputs and an output ICA dimension of 49. The first and eighth PCA components are both azimuthally symmetric and are removed from the analysis to increase contrast in the relevant final ICA outputs, described below.

The ICA components display several major attributes. Most striking, is the presence of distinct crystalline grains with three-fold rotationally symmetric diffraction patterns, shown in Fig. 6(a) (k-space components, top subpanels). The spatial distribution of each unique grain type is shown in the associated real space component (bottom subpanels).

The first noise component in Fig. 6(b), shows the mean response of the sample under the electron beam. The majority of the noise components appear as spatially homogeneous, illustrated by most of the bottom subpanels of Fig. 6(b), with no apparent physical meaning other than instrumentation noise. Finally, we find rare components in Fig. 6(b) that are not spatially homogenous and have hexagonal diffraction patterns with approximate two-fold rotational symmetry. We attribute these components to describing differences in tilt parallel to the beam.

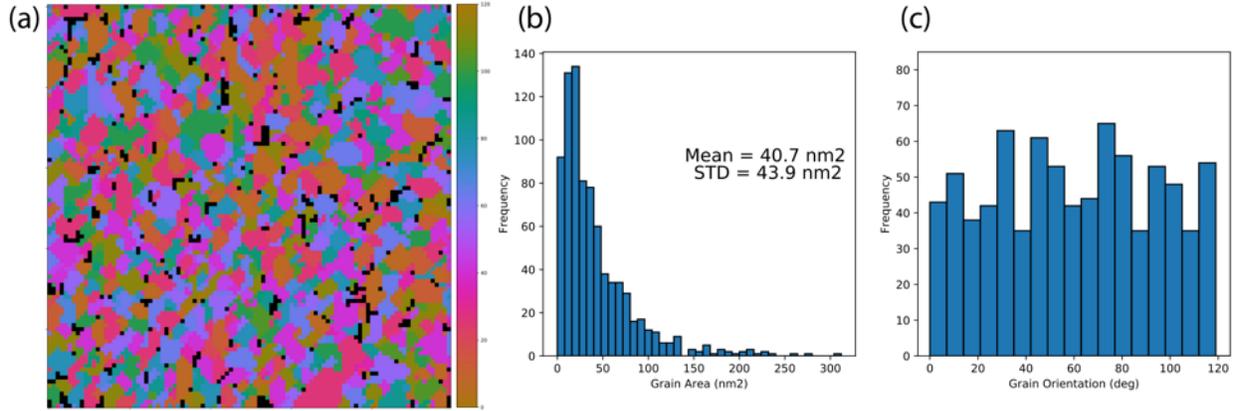

Fig. 7: Results of MVA grain size/orientation analysis for polycrystalline $WS_2$. The image in (a) shows the spatial distribution of distinct crystalline grains, with the color scale indicating the crystallographic orientation. The black regions are voids where no crystallographic component has appreciable value. The histogram in (b) shows the distribution of grain areas, peaking at 10 $nm^2$, and dropping to near zero at 150 $nm^2$. The grain orientation angle distribution is shown in (c). The orientation is defined by the angle of the most intense peak of the k-space component, modulo 120 degrees (due to the three-fold symmetry of the underlying 1H lattice). The distribution is approximately flat, indicating no preferred grain orientation.

The model that has emerged from ICA unmixing is that the sample is largely composed of distinct crystalline grains, each with three-fold rotational symmetry consistent with the 1H phase. Importantly, this three-fold symmetry is not readily apparent from the raw data and has only emerged after MVA processing.

To extract the details of the grain size and rotational orientation across the $WS_2$ specimen, we employ image featurization (using Hu image moments[29]) and a clustering algorithm (affinity propagation[30]) to automatically sort the components into groups that have similar spatial features and spectral symmetry. This clustering analysis is applied to the featurized IC diffraction patterns, and the same grouping is then applied to the corresponding IC spatial images. We use standard thresholding and particle analysis methods to generate histograms of the grain sizes and orientations (Fig. 7)(b-c) and a rotational orientation grain map (Fig. 7)(a). From the grain area distribution we see that the CVD synthesis method used to produce the WS$_2$ films favors small grains (~10 nm$^2$), but larger ones are also present up to approximately 100 nm$^2$. We also find from the diffraction patterns that there is no preferred orientation for the crystalline grains.

In order to assess the efficacy of our MVA methodology, we compare the results using MVA to similar results obtained using traditional analysis methods[31]. Fig. 8 shows both results side by side. The range of angles for the grain map using MVA has been reduced to 60 degrees for a fair comparison (hence Figs. 7(a) and 8(b) are very similar, but not identical). We see that although the agreement is not perfect, the two maps have a high level of similarity, confirming that our MVA methodology is a useful tool for quickly assessing the approximate distribution of sample parameters in a high dimensional data set.

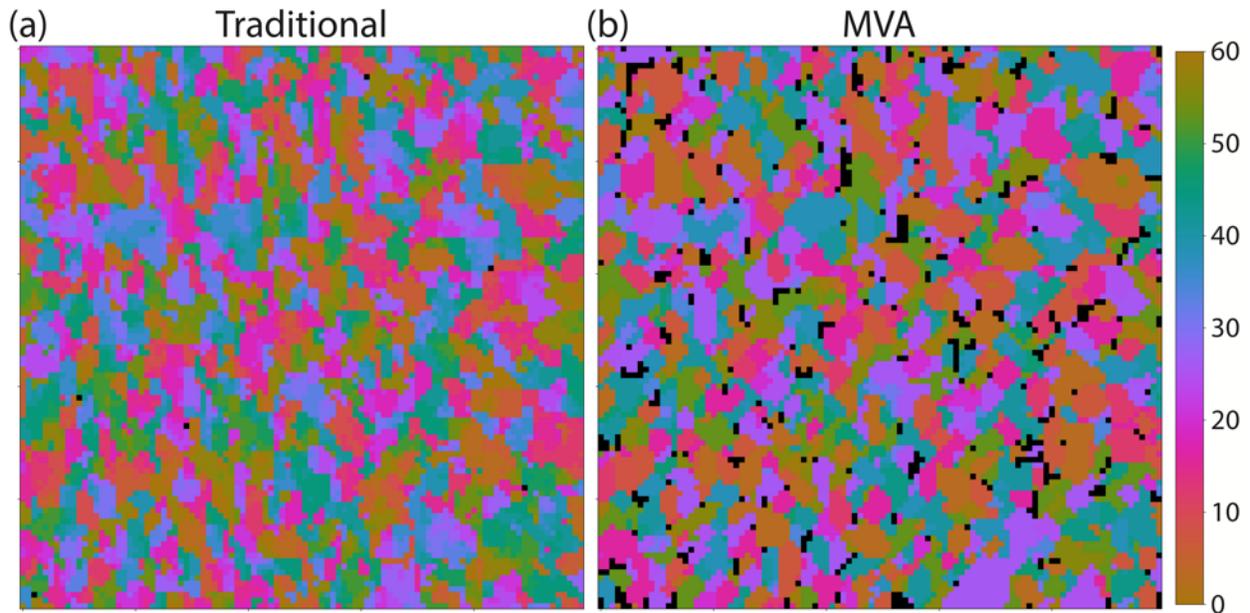

Fig. 8: Grain size/orientation map from polycrystalline WS$_2$. Panel (a) shows the results using a traditional analysis and panel (b) shows the results using MVA methodology.

## Methods

### Samples

Polycrystalline WS2 films are obtained via WO3 conversion[13]. 2 nm of WO3 is deposited using ALD onto 30 nm thick electron transparent Si3N4 TEM windows. The WO3 is then converted to WS2 by flowing H2S gas over the films at elevated temperature.

### 4DSTEM Imaging

4DSTEM data are collected using an FEI Titan 80-300 operated at an acceleration voltage of 200 kV. Electron diffraction patterns are acquired at 30 frames per second using a 14 bit Gatan Orius 830 CCD Camera. The electron probe has a diameter of 2 nm with a convergence angle 0.6 mrad and dose rate of 109 $e^-/Å^2$ sec. The map is acquired over a square 100 spatial pixels per side with a step size of 2 nm between pixels.

### Clustering Analysis

For the input of the clustering analysis we first binarize each diffraction pattern using the Otsu method. Then we use the Hu image moments of the binarized diffraction patterns as the feature vector for clustering analysis. Hu moments are invariant under translation, scale and rotation and therefore sort images into groups with similar symmetry[29]. The feature vectors are given to scikit learn's affinity propagation algorithm which finds the optimal number of clusters and assigns each input into a cluster[30,32].

### Particle Analysis

We use the IC images shown in Fig. 4a as the input for the analysis. First the images are thresholded using the Otsu method. We then perform a binary closing operation to remove grains smaller than two pixels and close holes within any given grain. Finally, we calculate the region properties of each grain using functions available in scikit image's measure package[33].

**Author Contributions**

BS, AZ, and SA conceived the project. BS performed the electron microscopy and analyzed the data. CTC, CK, TK, AS, and SA developed the methodology for fabrication of electron transparent, polycrystalline TMD films and prepared the $WS_2$ samples. BS and AZ wrote the manuscript. SA and AZ supervised the project. All authors proofread the paper.

**Competing Interests**

The authors declare no competing interests.

**Data Availability**

The data presented in the manuscript and analysis code that were used in this study are available from the corresponding authors upon reasonable request.

**Acknowledgments**

This work was supported by the Director, Office of Science, Office of Basic Energy Sciences, Materials Sciences and Engineering Division, of the U.S. Department of Energy under Contract No. DE-AC02-05-CH11231, with support coming primarily from the sp2-Bonded Materials Program (KC2207), which provided for development of the theoretical framework and 4DSTEM experiments, and additionally from the van der Waals Heterostructure Program (KCWF16) which provided for preliminary sample preparation and characterization. This work was additionally supported by the National Science Foundation under Grant # DMR-1807233 which provided for conventional ADF TEM imaging. Work at the Molecular Foundry was supported by the Office of Science, Office of Basic Energy Sciences, of the US Department of Energy under Contract No. DE-AC02-05CH11231. We acknowledge Christoph Gammer, Colin Ophus, and Peter Ercius from the NCEM facility at the Molecular Foundry for developing the code used to acquire the 4DSTEM datasets and for helpful discussion.

11. Kang, K. *et al.* High-mobility three-atom-thick semiconducting films with wafer-scale homogeneity. *Nature* **520**, 656–660 (2015).

12. Song, J.-G. *et al.* Controllable synthesis of molybdenum tungsten disulfide alloy for vertically composition-controlled multilayer. *Nat. Commun.* **6**, 7817 (2015).

13. Kastl, C. *et al.* The important role of water in growth of monolayer transition metal dichalcogenides. *2D Mater.* **4**, 021024 (2017).

14. Kim, K. *et al.* Grain Boundary Mapping in Polycrystalline Graphene. *ACS Nano* **5**, 2142–2146 (2011).

15. Huang, P. Y. *et al.* Grains and grain boundaries in single-layer graphene atomic patchwork quilts. *Nature* **469**, 389–392 (2011).

16. Belianinov, A. *et al.* Identification of phases, symmetries and defects through local crystallography. *Nat. Commun.* **6**, 7801 (2015).

17. Belianinov, A., Kalinin, S. V. & Jesse, S. Complete information acquisition in dynamic force microscopy. *Nat. Commun.* **6**, 6550 (2015).

18. Jesse, S. *et al.* Big Data Analytics for Scanning Transmission Electron Microscopy Ptychography. *Sci. Rep.* **6**, 1–8 (2016).

19. Belianinov, A. *et al.* Big data and deep data in scanning and electron microscopies: deriving functionality from multidimensional data sets. *Adv. Struct. Chem. Imaging* **1**, 6 (2015).

20. Sarahan, M. C., Chi, M., Masiel, D. J. & Browning, N. D. Point defect characterization in HAADF-STEM images using multivariate statistical analysis. *Ultramicroscopy* **111**, 251–257 (2011).

21. Chen, Z. *et al.* Practical aspects of diffractive imaging using an atomic-scale coherent electron probe. *Ultramicroscopy* **169**, 107–121 (2016).